# Nanostructured p-p(v) junctions obtained by G-doping

A. Tavkhelidze [1] *, L. Jangidze [2] and G. Skhiladze [2]

[1] Ilia State University, Cholokashvili Ave. 3-5, 0162 Tbilisi, Georgia
[2] Institute of Micro and Nano Electronics, Chavchavadze Ave. 13, 0179 Tbilisi, Georgia

E-mail: avtotav@gmail.com

**Abstract**

Recently, geometry-induced quantum effects in periodic Si nanostructures were introduced and observed. Nanograting has been shown to originate geometry induced doping or G-doping. G-doping is based on reduction of density of quantum states and is n-type. Here, fabrication and characterization of G-doping based compensated p-p(v) junctions is reported. The p-p(v) abbreviation is introduced to emphasize voltage dependence of G-doping level. The p-type Si wafer is used for sample fabrication. First, two square islands are shaped at the surface of the wafer. Next, entire Si surface is oxidized to grow thin SiO2 insulating layer. Two windows are opened in SiO2 and nanograting is fabricated inside one of the windows using laser interference lithography followed by reactive ion etching. G-doping p-p(v) junction between p-type substrate and the nanograting layer is formed. Next, metal contacts are deposited on both islands to measure electrical characteristics of p-p(v) junction. To exclude contact resistance input Van der Pauw method is used. Obtained I-V curves are diode type with extremely low voltage drop in forward direction and reduced reverse current. Experimental I-V curves are fitted to the Shockley equation and ideality factor is found to be in the range of 0.2-0.14. Such a low values confirm that p-p(v) junction is fundamentally different from the conventional p-n junction. This difference is ascribed to G-doping which, unlike conventional doping, is external voltage dependent.

## 1. Introduction

Present developments in interference lithography enabled fabrication of periodic nanostructures [1, 2]. Semiconductor nanograting (NG) layers have been introduced and fabricated [3,4]. The periodic structure, have been shown to dramatically change the electronic [4], thermoelectric [5], optical [6] and electron emission [7] properties when the nanograting dimensions are low enough to approach the de Broglie wavelength of electrons. This is due to the extraordinary boundary conditions imposed by a NG on the electron wave function. They make forbidden definite quantum states [8], and density of quantum states reduce. Electrons rejected from the valence band occupy empty quantum states in the conduction band. The electron concentration in the conduction band enhance, which is named as geometry-induced electron doping or G-doping [3]. It is equal to donor doping from the point of view of the raise in $n$ and Fermi energy. However, there are no ionized impurities.

The NG geometry belongs to a class of nonintegrable quantum systems (quantum billiards) which are widely investigated [9, 10].

There are some experimental results demonstrating nanostructure induced doping. In disordered nanostructures obtained by wet etching of p-Si [11] both n-type and p-type doping was observed. Periodic nanostructures made by laser radiation interaction with surfaces of Si, Ge and SiGe crystals demonstrate n-type doping [12]. In mesoporous p-Si charge carriers disappear [13]. G-doping mechanism can be employed to explain these results.

Thin Si nanograting layers have been studied experimentally. It was found, that resistivity and Hall voltage have metal type temperature dependences [4]. Dielectric function determined by ellipsometry show metallization of the layer [6]. Intensive photoluminescence was observed in Si in spite of indirect band gap [14].

In this work, we report on fabrication and investigation of G-doped p-p(v) junctions.

## 2. Sample Preparation and Characterization

Silicon wafers p-type were used for sample preparation. Wafers were p-type (Boron) with resistivity 1-10 Ω x cm and thickness 200 microns. Samples were 10mm x20mm chips with NG and plain islands and corresponding metal contacts. On the reference plain side NG was not formed. Figure 1 shows simplified process flow (only the NG side is shown).

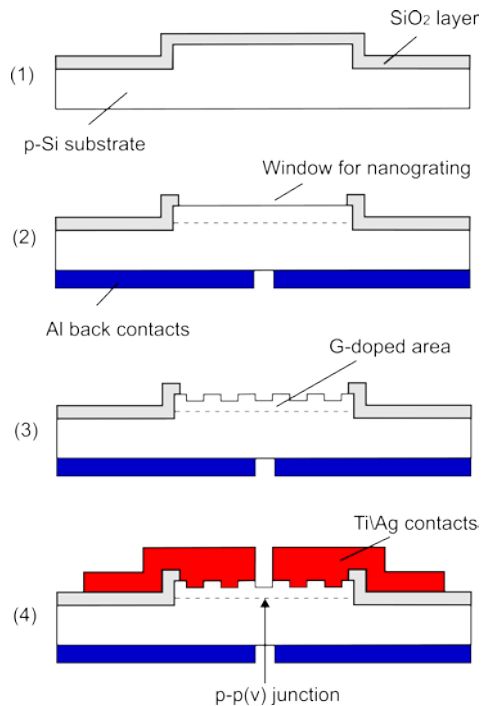

**Figure 1**. The schematically the process flow for G-doped solar cell preparation:

The G-doped solar cells were prepared in the following steps:
1. Back contact p+ type was made by Boron diffusion at 1050 $^0$C – 10 minutes. Diffusion depth was 0.7-0.8 μm. Next, square island of 2.2mm x 2.2mm and height of 0.2-0.4 μm was formed on the front side of the wafer using photolithography. Next, SiO$_2$ insulator layer was grown using wet thermal oxidation 1000 $^0$C during 10 minutes. Thickness of SiO$_2$ layer was 150-170 nm. Next, 0.6 μm Al contact was deposited on the back side and backed at 540 $^0$C during 12 minutes.
2. The 2mmx2mm window was opened in SiO$_2$ layer using photolithography and wet etching of SiO$_2$.
3. The NG with line width of 150 nm was fabricated inside the window using laser interference lithography [4] and subsequent reactive ion etching. First, negative photo resist ma-N 2401 was applied to the sample. Next, laser interference lithography was done using Blue-Violet (375 nm) semiconductor laser as a coherent light source. The 25-35 nm depth NG was made using reactive ion etching in CF$_4$.
4. Contacts were made by Al thermal evaporation and Ti\Ag magnetron sputtering. Conventional lift-off technology was used in both cases. The Ti\Ag films were grown using DC magnetron sputtering during the single vacuum process at a substrate temperature of 250 °C [4]. Ti\Ag contacts were not backed.

Current-voltage (I-V) curves were recorded using both the 4-probe Van der Pauw method and 2-probe method. Using the 2-probe method was necessary, as the 4-probe method alone does not provide information about contacts. A Keysight multimeters 34410A and 34410 were used to record 4-probe and 2-probe voltages, and Keysight E3647A current source was used to apply current. To exclude input of thermo powers voltage shifts (typically <10$^{-4}$ V) were compensated in the process of I-V curve building.

## 3. Results and discussion

Typical I-V curve of the p-p(v) junction (blue) is shown in Fig. 2. In the same figure I-V curve of reference plain junction is plotted. It is evident that NG produce diode type I-V curve and reference plain is close to ohmic contact. Forward voltage is very low with respect to Si diode and forward currents are quite high. The p-p(v) junction is formed thanks to reduction of quantum state density near the NG. Effective density of states ($N_V$ and $N_C$) in both valence and conduction bands reduce G times, where G is geometry factor [3]. Consequently, internal carrier concentration near the NG also reduces

$$n_i^{NG} = n_i / G \qquad (1)$$

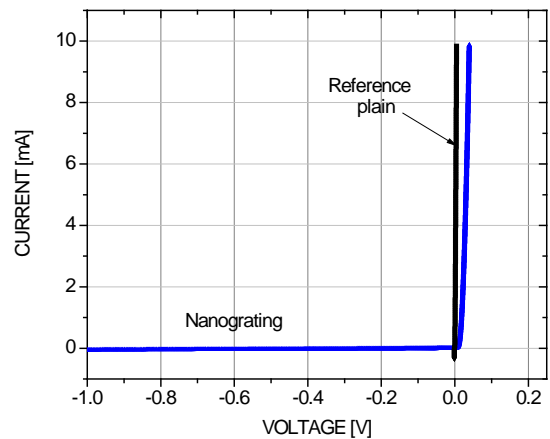

**Figure 2.** Typical I-V characteristics of NG and reference plain (sample M154).

(follows from the definition of $n_i$). However, mass action law rewritten as

$$p_{NG} n_{NG} = n_i^2 / G^2 \qquad (2)$$

should be satisfied. This leads to reduction of hole concentration near the nanograting $p_{NG} < p$. In our opinion, such a reduction of hole concentration is the reason for p-p(v) junction formation.

The same I-V curve (nanograting in Fig.2), normalized and plotted in semilog scale is shown in Fig. 3. In both forward and reverse directions I-V curve is diode type. In forward direction current increases very rapidly following Shockley equation

$$J / J_0 = \exp(eV / \eta KT) - 1 \qquad (3)$$

with ideality factor $\eta = 0.14$. Here, $e$ is electron charge; V is applied voltage K is Boltzman constant; T is absolute temperature. Fig. 3 clearly indicates that p-p(v) junction is fundamentally different from p-n junction (where $\eta = 1 - 2$). Forward current increases much faster than it would be even for the ideal p-n junction. This can be explained by voltage dependence of $G = G(V)$. Such dependence follows from the voltage dependence of the p(v) layer thickness (layer in close proximity to the NG with reduced quantum state density).

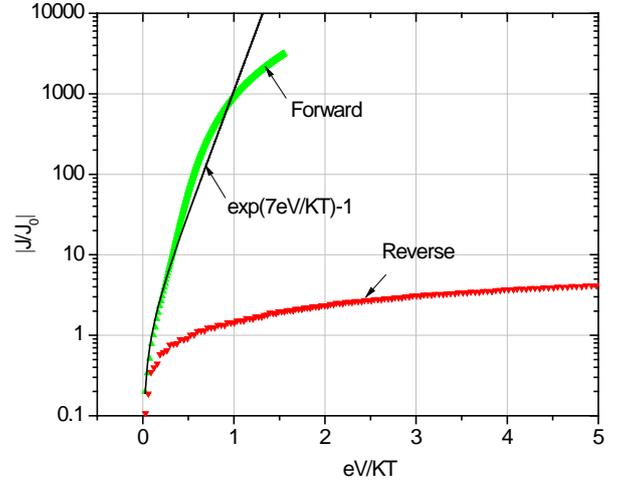

**Figure 3**. The I-V characteristics of p-p(v) junction (sample M154). The value $\eta = 0.14$ was used for normalization.

Physical mechanism responsible for such dependence is unclear yet. One of the explanations is charge accumulation layer thickness dependence on applied voltage.

Forward voltages are in the range 20-60 mV and are much lower than for silicon p-n junctions and even considerably lower than for Si Schottky junctions [17]. The reverse current increases with voltage faster than it would be in the case of ideal p-n junction, but considerably slower than it would be for the real silicon p-n junction [16, p. 97].

Saturation current density, forward and reverse currents for 9 samples are given it Table 1.

**Table 1**. Parameters of p-p(v) junctions. Last column indicates how much p-p(v) junction reverse current is reduced with respect to silicon p-n junction. Value of reverse $(J/J_0)_{p-n} = 1000$ for $eV/KT = 30$ is taken from [16, p. 97].

| Sample # | $\eta$ | $J_0$ [μA/cm²] | Forward $J/J_0$ at $\alpha^* = 1$ | Reverse $J/J_0$ at $\alpha = 1$ | Reverse $J/J_0$ at $\alpha = 10$ | Reverse $J/J_0$ at $\alpha = 30$ | Reverse $\frac{(J/J_0)_{p-p(v)}}{(J/J_0)_{p-n}}$ at $\alpha = 30$ |
|---|---|---|---|---|---|---|---|
| M153 | 0.2 | 15 | 100 | 2 | 15 | 35 | 0.035 |
| M154 | 0.14 | 77 | 1000 | 2 | 7 | 15 | 0.015 |
| M157 | 0.2 | 26 | 250 | 4 | 20 | 40 | 0.04 |
| M159 | 0.17 | 7.5 | 900 | 5 | 20 | - | - |
| M160 | 0.2 | 27 | 100 | 5 | 30 | 70 | 0.07 |
| M164 | 0.18 | 8.2 | 200 | 2 | 15 | - | - |
| M166 | 0.17 | 17 | 300 | 2 | 7 | 10 | 0.01 |
| M168 | 0.17 | 6.9 | 400 | 7 | 15 | 20 | 0.02 |
| M174 | 0.17 | 23 | 2000 | 3 | 10 | 20 | 0.02 |

* $\alpha \equiv eV / KT$

Ideality factor is quite low and keeps inside the relatively narrow range $\eta$ = 0.14-0.2. Saturation current density $J_0$ varies from sample to sample. The same is true for forward current, while reverse currents do not vary much. Variation of saturation current can be ascribed to the variation of carrier concentration in p-Si (1-10 Ω x cm) wafers. Large variation of normalized forward current $J/J_0$ can be ascribed to the exponential dependence equation (3). Relative stability of reverse currents can be ascribed to saturation which also follows from equation (3). Consequently, we find good overall matching of experimental data to the equation (3). However, ideality factor is less than 1 and is very low (in the range $\eta$ = 0.14-0.2). Our explanation is dependence of G-doping level on applied voltage.

Saturation current density $J_0$ is of order of $10^{-5}$ A cm$^{-2}$ which is many orders of magnitude higher with respect to silicon p-n junction (typically $10^{-10}$ A cm$^{-2}$). Our explanation is that in p-p(v) type junction $J_0$ is determined by majority carrier density instead of minority carrier density (as for p-n junction).

The reverse current voltage dependence for two samples is show in Fig. 4. Reverse currents increase smoothly at least until 4V reverse bias and are much less with respects to silicon p-n junction, resulting in $(J/J_0)_{p-p(v)} / (J/J_0)_{p-n} << 1$ for the value $eV/KT = 30$ (Table 1). Value of reverse $(J/J_0)_{p-n} = 1000$ for $eV/KT = 30$ is found in [16, p. 97]. Our explanation is following. In the case of G-doping, we have reduced value of $n_i^{NG}$ equation (1). However, diffusion current which is defined by majority carrier concentration (instead of minority for p-n junction) still dominates over generation current. It happens despite Si material and relatively low temperature ($T$=300K). Consequently, reverse current is less dependent on carrier generation and normalized reverse current is much less with respect to silicon p-n junction.

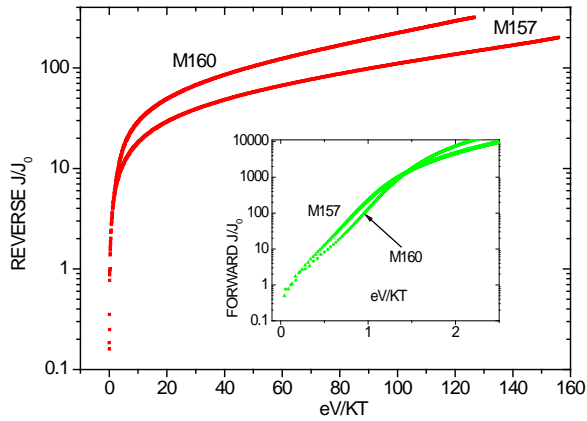

**Figure 4**. Reverse current voltage dependences (red) for samples M157 and M160. Corresponding forward current dependences (green) are given in the insert.

Further experiments are required to improve ohmic contacts, record C-V curves and measure high frequency characteristics.

**Conclusions**

We fabricate and characterize G-doping based p-p(v) junctions. The p-type Si wafer (1-10 Ω x cm) was used for sample fabrication. G-doped p-p(v) junction has been formed between the p-type substrate and nasnograting with line width of 80-150 nm and depth of 25-35 nm. Measured I-V curves were diode type with extremely low voltage drop in forward direction and reduced reverse currents. Experimental I-V curves were fitted to the Shockley equation and ideality factor was found to be in the range of 0.2-0.14. Such a low values of ideality factor confirm that p-p(v) junction is fundamentally different from the conventional p-n junction. We attribute this difference to G-doping which, unlike conventional doping, is voltage dependent. Data collected from 9 samples indicate that saturation currents are several magnitudes higher with respect to silicon p-n junction. This is explained by key contribution of majority carriers (holes) is saturation current. Reverse currents are 1-2 order of magnitude lower with respect to silicon p-n junction which is explained by leading role of diffusion current. Forward voltage drops are 30-60 mV and are considerably lower with respect to Schottky diode. Normalized reverse currents are much less with respect to p-n junction (at least for reverse bias up to 4V). The p-p(v) diode has certain advantages over p-n and Schottky diodes and can find wide applications in power electronics and ultra high frequency electronics as well as in conventional electronics and solar cells.

**Acknowledgments**

We thank E. A. Katz, N. Gorj, I. Shah and Z. Taliashvili, for discussion and support. The authors thank the SRNSF (MTCU/91/3-250/15) and STCU (project 6191) for providing funding.